\documentclass[aps,prb,showpacs,twocolumn,groupedaddress]{revtex4}
\usepackage{graphicx}
\usepackage{dcolumn}
\usepackage{bm}
\usepackage{amsmath}
\usepackage{amssymb}

\newcommand{\beq}    {\begin{equation}}
\newcommand{\enq}    {\end{equation}}
\newcommand{\ceq}[1] {(\ref{#1})}

\newcommand{\rr}     {{\bf r}}

\newcommand{\EE}     {{\bf E}}

\newcommand{\seff}   {\sigma_{\rm EMT}}

\newcommand{\nav}    {\langle n\rangle}

\begin{document}

\title{Effective medium theory for disordered two-dimensional graphene}

\author{Enrico Rossi, S. Adam, S. Das Sarma}

\affiliation{Condensed Matter Theory Center, Department of Physics,
             University of Maryland, College Park, Maryland 20742, USA}

\date{\today}


\begin{abstract}

We develop an Effective Medium Theory to study the electrical
transport properties of disordered graphene.  The theory includes
non-linear screening and exchange-correlation effects
allowing us to consider experimentally
relevant strengths of the Coulomb interaction.  Assuming random
Coulomb impurities, we calculate the electrical conductivity as a
function of gate voltage describing quantitatively the full cross-over
from the fluctuations dominated regime around the Dirac point to the
large doping regime at high gate voltages. We find that the 
conductivity at the Dirac point is strongly affected by 
exchange correlation effects.

\end{abstract}


\maketitle


\section{Introduction}
Since the experimental realization of graphene 
\cite{novoselov2004, novoselov2005,zhang2005} 
the problem of understanding the electronic conductivity of graphene
has generated a large experimental and theoretical effort, given its
fundamental importance and its technological relevance.  The unusual
transport properties of graphene \cite{novoselov2004} arise mostly
from its ``Dirac spectrum'' -- a linear, zero-gap dispersion relation
of chiral Fermion carriers (electrons and holes), with the charge
neutral ``Dirac point'' defined as the singular point (with vanishing
density of states) where the electron and hole bands touch.  In this
work we present a theory that is able to describe at a
qualitative and semi-quantitative level most of the current bulk
transport experimental results on exfoliated graphene.  The theory is
designed specifically to handle the disorder induced density
inhomogeneities present in graphene and dominating the physics at the
Dirac point where the average carrier density is zero.  The theory is
general: it can be applied to systems other than graphene (as
formulated, it applies readily to graphene bilayers); it does not
depend directly on the source of the spatial inhomogeneities; and it can be
used to calculate other graphene transport properties such as
thermo-transport \cite{hwang2009} and magnetotransport.
The theory takes into account non-linear screening
effects and exchange-correlation terms via a Hartree-Fock
Local Density Approximation (LDA); in this sense the theory
is non-perturbative and therefore, within the limits of 
this approximation, is valid for large values of the graphene's fine structure
constant $r_s\equiv e^2/\hbar v_F\epsilon$, where $v_F$ is the Fermi velocity 
and $\epsilon$ the dielectric constant.
For graphene charge transport, the theory is powerful being able to:
1) Predict values of conductivity at the Dirac point, $\sigma_{min}$,
 in very good agreement with experiments;
2) Describe quantitatively the dependence of $\sigma$ on the
 doping $\propto V_g$ including the
 {\em crossover} or {\em plateau} region, i.e. the region where $V_g$
 is finite but smaller than the range of values for
 which $\sigma(V_g)$ exhibits a linear behavior;
3) Show that many-body terms are essential to understand
   the transport properties of graphene close to the Dirac point;
4) Explain the observed dependence of $\sigma_{min}$ on the
 sample quality;
5) Explain the dependence of $\sigma_{min}$ on
 $r_s$ observed experimentally \cite{jang2008} and
 predict its behavior over a wide range of $r_s$,
 a prediction that could be tested experimentally;
6) Show that at large doping the spatial density fluctuations
 do not modify the linear dependence of $\sigma$ on $V_g$; and
7) Predict the transport properties at very low but finite
long-range disorder; a regime not accessible in previous theories.

The success of the transport theory presented in this work relies on
the proper treatment of the effects of the spatial carrier
inhomogeneity.  In a metal or semiconductor, defects play a double
role \cite{landauer1952}: they act as scattering centers and locally
modify the conduction-band carrier density.  A defect effectively
shifts the local Fermi level away from its average value, which has
dramatic consequences on the properties of graphene close to the Dirac
point where the average density, $\langle n\rangle$, is zero and
therefore the spatial density fluctuations completely dominate the
physics.  The source of the disorder \cite{neto2009} can be of
different nature: short-range scatterers, such as atomic defects in
the graphene lattice structure, ripples \cite{meyer2007} or long range
scatterers such as charged impurities, but all will induce spatial
inhomogeneities of the carrier density that close to the Dirac point
result in the formation of electron-hole, {\em e-h}, puddles
\cite{hwang2007, adam2007, martin2008, rossi2008}.
This situation has emerged as the one realized in current experiments
on exfoliated graphene. The ideal case of zero average density
and zero amplitude of the density fluctuations has been considered
in previous works that predicted  $\sigma_{min}$
to be either 0 or $\infty$, depending on the nature of the disorder,
or, in the limit of vanishing disorder, 
finite and universal \cite{fradkin1986,ludwig1994,katsnelson2006,
fritz2008,kashuba2008,bardarson2007,nomura2007}.

The formalism that we
present does not depend on the source of the inhomogeneities, however
recent experimental results~\cite{tan2007,chen2008,jang2008} provide
convincing evidence that random charged impurities -- located in the
graphene environment -- are the dominant source of disorder in
graphene. Therefore for concreteness we assume the carriers in
graphene to be subject to a disorder potential due to a 2D random
distribution of uncorrelated Coulomb impurities with surface density
$n_{imp}$, placed at a distance $d$ from the graphene layer.
The paper is organized as follows: in section \ref{sec_model}
we develop and justify the effective medium theory for graphene
in presence of a 2D random distribution of charged impurities,
in section \ref{sec_results} we present the results for the
conductivity of disordered graphene and finally in section \ref{sec_conclusions}
we summarize the main results and discuss the range of validity
of the approach presented.


\section{Theoretical Model}
\label{sec_model}
As said in the introduction, in the remainder of the paper we
only consider the disorder induced by random charged impurities in
the graphene environment, motivated by the strong experimental
evidence \cite{chen2008, tan2007} that this type of disorder
dominates in current experiments on exfoliated graphene on 
${\rm SiO_2}$ substrate.
In order to be able to correctly describe the properties
of graphene close to the Dirac point an accurate characterization
of the strong, disorder induced, density fluctuations is necessary.
This requires the calculation of the ground state for 
interacting massless Dirac electron in presence of disorder,
a very challenging problem. In absence of disorder one can use
a Kohn-Sham-Dirac Density-Functional-Theory, DFT,  that appropriately takes into account
the fact that in graphene the low energy quasiparticles
behave as chiral massless Dirac fermions \cite{polini2008}.
In this approach many-body effects enter via a local-density
approximation  exchange-correlation potential.
In presence of disorder the DFT-LDA approach is computationally
very expensive and as a consequence can only be used to
calculate the ground state for few disorder realizations \cite{polini2008} and
does not allow for a statistical characterization of the carrier density fluctuations
that requires the calculation of the disorder averaged quantities.
A computationally cheaper approach is the 
microscopic Thomas-Fermi-Dirac, TFD, theory that recently some
of us have developed and used to study the electronic
structure of disordered graphene \cite{rossi2008}.
The TFD approach is very similar to the DFT-LDA one,
the difference being that in the TFD theory also
the kinetic energy is approximated by a local-density
functional, whereas in the DFT-LDA the kinetic term
is a differential operator that operates on the Kohn-Sham orbitals.
For single disorder realizations the TFD results can
be directly compared to the DFT-LDA results and we found
them to be in very good agreement for impurity density
values, $n_{imp}$, relevant for current experiments
on exfoliated graphene: $n_{imp}>10^9{\rm cm}^{-2}$.
The Thomas-Fermi theory in general is accurate when the ground-state density
varies on length scales bigger than the the Fermi
wavelength, $\lambda_F$, i.e.:
\begin{equation}
 \left[\frac{\nabla n}{n}\right]^{-1}\gg \lambda_F
 \label{eq:ineq1}
\end{equation}
Because $k_F =\sqrt{\pi n}$ inequality \ceq{eq:ineq1} can be rewritten as:
\begin{equation}
 \left[\frac{\nabla n}{n}\right]^{-1}\gg \frac{1}{\sqrt{\pi n}}.
 \label{eq:ineq2}
\end{equation}
Inequality \ceq{eq:ineq2} is equivalent to say that
in any region with homogeneous carrier density the number
of carriers has to be bigger than 1. As shown below
the non-linear screening properties of graphene
guarantee that in presence of charged impurities this condition
is satisfied for graphene also close to the Dirac point.

To summarize the main properties of the graphene carrier distribution
close to the Dirac point in the lower color
plot of Fig.~\ref{fig:5} we show a typical result obtained using the TFD theory
for a given disorder realization with $n_{imp}=5\times 10^{11}$~cm$^{-2}$,
$d=1$~nm and $r_s=0.8$, values that are typical for current
exfoliated graphene samples and accurately estimated
by comparing the Boltzmann theory and the experimental
results at high doping \cite{tan2007}.
This plot exemplifies the main qualitative properties of
the carrier distribution close to the Dirac point:
\begin{figure}[ht]
 \begin{center}
  \includegraphics[width=8.00cm]{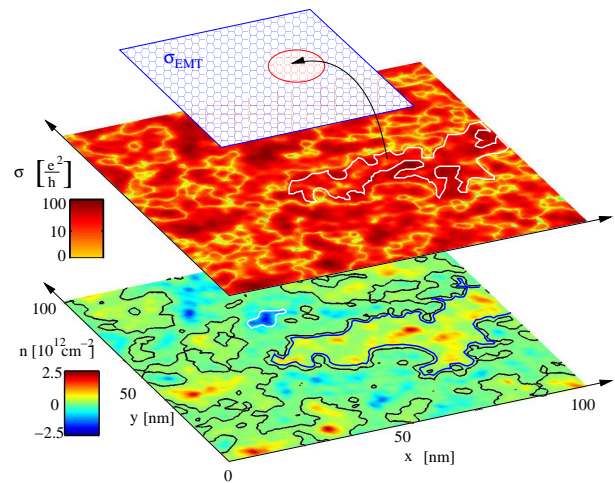}
  \caption{
           (Color online).
           $n(\rr)$ lower plot, and corresponding $\sigma(\rr)$, middle plot,
	   at the Dirac point for a single 
           disorder realization: $n_{imp}=5\times 10^{11}$~cm$^{-2}$, $d=1$~nm $r_s=0.8$.
	   Top panel shows a schematic of the {\em Effective Medium} 
	   used to describe the strongly inhomogeneous state of graphene
	   close to the Dirac point.
          }
    \label{fig:5}
 \end{center}
\end{figure}
I)   the distribution is characterized by e-h puddles 
     in quantitative agreement
     with surface probe experiments~\cite{martin2008, zhang2009};
II)  the typical size of an e-h puddle, defined as
     a region with same-sign charges, is of the order of the sample
     size, $L$, as expected for a semimetal close to the neutrality point;
III) $n(\rr)$ is characterized by two types of inhomogeneities,
     namely, wide regions (i.e. big puddles spanning the system size) of
     low density (an example is shown in Fig 1 with the blue contour
     that contains $\approx 10$ electrons); and few narrow regions of
     high density (an example is shown in Fig 1 with the white contour
     containing $2$ electrons);
IV)  the narrow puddle regions (peaks, dips) have a correlation length $\xi$ that depends on
     $n_{\rm imp}$, $r_s$ and $d$. \cite{rossi2008} For the parameter values used 
     in Fig.~\ref{fig:5}, typical for current experiments, $\xi\approx 10$nm. 
The combination of the high density in the peaks/dips and the fact
that in the low density regions $n(\rr)$ varies over scales much
bigger than $\xi$ guarantees that the inequality \ceq{eq:ineq2}
is satisfied over the majority of the
graphene sample and therefore justifies the TFD theory.  The
properties I)-IV) are essential for the development and justification
of the effective medium theory, EMT, for graphene presented in this
work.

In presence of disorder, the problem of graphene transport then
becomes one of how to properly take into account the strong density
inhomogeneities shown in Fig.~\ref{fig:5}.  At high doping the
graphene carrier density can be assumed to be homogeneous.  In this
regime, $\sigma$ has been shown experimentally to scale linearly with
$\nav$.  
The theory that has been able to describe most successfully, away from the 
Dirac point, the large number of experimental results for graphene transport
is scattering from random charge impurities \cite{nomura2006,ando2006,cheianov2006,hwang2007}, where,
within the RPA-Boltzmann approximation, we have:
\begin{equation} 
 \sigma(\rr) = \frac{2 e^2}{h} \frac{|\nav|}{n_{\rm imp}} \frac{1}{F(r_s, d)}
 \label{eq:sigma_r}
\end{equation}
where $F(r_s, d)$ was given in Ref.~\onlinecite{adam2007}.  An important
conceptual step to derive a correct transport theory valid close to
the Dirac point is to introduce a local spatially varying ``puddle''
conductivity $\sigma(\rr)$. 
The local conductivity is a well defined
quantity as long as $\sigma(\rr)$ varies on length scales that are
bigger than the mean free path $l$ i.e.:
\beq
 \hspace{0.2cm} \left|\frac{\nabla\sigma(\rr)}{\sigma(\rr)}\right|^{-1}\gg l;
 \label{eq:ls}
\enq
Because, as shown in Fig.~\ref{fig:5},
close to the Dirac point the density profile is characterized by large
puddles with $n\neq 0$ the RPA-Boltzmann theory is expected to be
valid locally inside the e-h puddles. To estimate $\sigma$ in a puddle
we therefore use Eq.~\ceq{eq:sigma_r} replacing $\nav$ by its local
value $n(\rr)$ so that in Eq.~\ceq{eq:ls} we can replace
$\nabla\sigma(\rr)/\sigma(\rr)$ with $\nabla n(\rr)/n(\rr)$.
Considering that $l=h\sigma/(2e^2k_F)$, using Eq.~\ceq{eq:sigma_r},
for disordered graphene inequality \ceq{eq:ls} takes the form:
\beq
 \hspace{0.2cm} \left|\frac{\nabla n(\rr)}{n(\rr)}\right|^{-1} \gg
 \frac{1}{\sqrt{\pi}F(r_s,d)}\frac{\sqrt{n}}{n_{imp}}.
 \label{eq:ls2}
\enq
For graphene on ${\rm SiO_2}$
($r_s=0.8$) $1/F(r_s,d)\approx 10$.
Close to the Dirac point there are few narrow
regions of high density and size $\xi(d, n_{\rm imp}, r_s)$. For
$r_s=0.8$, $d=1$~nm, $n_{\rm imp}\sim 10^{11}$~cm$^{-2}$ 
(parameter values relevant for current experiments on 
exfoliated graphene) is $\xi\sim
10$~nm$\lesssim l$ as also shown in the lower panel of Fig.~\ref{fig:5}.  
However, the
regions where the carrier density changes over length scales of the
order of $\xi$ are very sparse and the density landscape is
characterized by wide regions 
where the density is approximately
constant 
on much bigger length scales, of the order of the system size.
Because the small puddles of size $\approx\xi$ are isolated
(i.e. do not form a path spanning the whole sample)
and occupy a small area fraction (we estimated it to be less than 20\%~\cite{rossi2008})
we can neglect their contribution to the graphene conductivity.
In addition, given their high density, steep
gradients at the boundaries, and the fact that $\xi\ll l$, 
the small puddles should provide very small resistance to the carriers' motion.
In the large regions where $n(\rr)$ is almost uniform 
$n$ is approximately equal to the root mean square of the density distribution
$n_{rms}\sim n_{imp}$ \cite{adam2007,rossi2008}, and as a consequence in these
regions the inequality \ceq{eq:ls} is satisfied for experiments on bulk
exfoliated graphene for which we have $L\sim 1\mu{\rm m}$ 
and $n_{imp}$ is in the range $[10^{11}-10^{12}]$~cm$^{-2}$.
We conclude that the non-linear
screening properties of graphene that justify the use of the TFD
theory, and the linear dependence of $\sigma$ on $n$
(Eq.~\ceq{eq:sigma_r}), allow for the definition of a local
conductivity for graphene at the Dirac point.
The conductivity spatial profile corresponding to the density profile
of the lower color plot of Fig.~\ref{fig:5}, is shown in the upper
color plot of the same figure.

Close to the Dirac point graphene can then be thought of
as a highly heterogeneous material for which the EMT should be
appropriate to calculate its transport properties. Even though the EMT
was proposed long ago for heterogeneous materials \cite{bruggeman1935,
landauer1952}, the development and {\em justification} for graphene is
not obvious and is the central result of this work.  To derive the EMT
one considers a single homogeneous region (e.g. area encircled by the
white perimeter in the color plot for $\sigma$ in Fig.~\ref{fig:5})
with conductivity $\sigma$ embedded in a uniform medium with effective
conductivity, $\sigma_{\rm EMT}$ and then requires the spatially
integrated inhomogeneities of the electric field $\EE$, due to the
inclusion of the region with $\sigma\neq\seff$, to be equal to zero.
For a 2D system this requirement translates into the equation:
\beq
 \int d^2 r\frac{\sigma(\rr)-\sigma_{EMT}}{\sigma(\rr) + \sigma_{EMT}} = 0.
 \label{eq:s2}
\enq
The EMT is valid if the conductance across the regions where
$\sigma$ is homogeneous is higher than the conductance inside
these regions. For graphene at the Dirac point this requires
that  the conductance across the
disorder induced p-n junctions, PNJ, (shown as black lines in the
lower color-plot of Fig.~\ref{fig:5}) is higher than the conductance
inside the puddles.  
Close to a PNJ Eq.~\ceq{eq:sigma_r} is not valid.  If $D$ is the
length scale over which $n$ changes across a PNJ and $k_F=\sqrt{\pi n}$
is the Fermi wave vector at the sides of the PNJ, then using TFD results
for $n(\rr)$ we find $k_F D\sim 1$; for single disorder realizations
this result can also be inferred from the DFT-LDA calculations \cite{polini2008}.  
In this limit, the conductance per unit length of a PNJ,$g_{\rm PNJ}$  
is given by \cite{cheianov2006b, zhang2008}:
\beq
 g_{\rm PNJ}=\frac{e^2}{h}k_F.
 \label{eq:gpnj}
\enq
Let $p$ be the perimeter of a typical puddle and $\Gamma$ 
its form factor then the condition that
the conductance across the PNJ must be bigger than
the conductance of a puddle takes the form:
\beq
 \frac{e^2}{h}\sqrt{\pi n} p\gg \Gamma \frac{2 e^2}{h} \frac{|n|}{n_{\rm imp}} \frac{1}{F(r_s, d)}.
\enq
i.e.
\beq
 p \gg \frac{2\Gamma}{\sqrt{\pi}F(r_s,d)}\frac{\sqrt{n}}{n_{imp}}.
 \label{eq:ineq3}
\enq
where $n$ is the density at
the sides of the PNJ: $n\sim n_{rms}\sim n_{imp}$ \cite{adam2007,rossi2008}.
Given the macroscopic size of the
majority of the e-h puddles the length of the PNJ, $p$, is of the order of
the sample size $L\sim 1\mu{\rm m}$, and then the inequality
\ceq{eq:ineq3} is satisfied.  We can therefore conclude
that the EMT developed here properly describes the electronic
transport properties of bulk disordered graphene samples and that
close to the Dirac point, transport in bulk graphene is dominated by
scattering processes {\em inside} the puddles and not {\em across}
their boundaries as was assumed in previous works \cite{cheianov2007}.
Disorder averaging Eq.~\ceq{eq:s2} we find:
\beq
 \left\langle\int d^2 r\frac{\sigma(\rr)-\sigma_{EMT}}{\sigma(\rr) + \sigma_{EMT}} = 0\right\rangle 
 \Longleftrightarrow
 \int d \sigma\frac{\sigma-\sigma_{EMT}}{\sigma + \sigma_{EMT}}P(\sigma) = 0,
 \label{eq:s4}
\enq
where $P(\sigma)$ is the probability for the local value of $\sigma$. 
Using Eq.~\ceq{eq:sigma_r} we have:
\begin{equation}
 \int dn\frac{\sigma(n)-\sigma_{\rm EMT}}{\sigma(n)+\sigma_{\rm EMT}}P[n] = 0.
 \label{eq:semt}
\end{equation}

To obtain an accurate value of $\seff$ it is crucial to select the
correct form of $P[n]$.  Choosing $P[n]$ to be a Gaussian distribution
of width $n_{\rm rms}$, for $\sigma_{min}$ we find $\sigma_{\rm
EMT}^G\approx (e^2/h)[1.1460/F(r_s,d)][n_{rms}/n_{imp}]$.  Using a
Lorentzian approximation for $P[n]$ with width $n_{L}$, we find
$\sigma_{\rm EMT}^L = (e^2/h)[2/F(r_s,d)](n_{L})/(n_{\rm imp})$.
These analytical results allow for a direct comparison with other
results for $\sigma_{min}$ in the literature~\cite{adam2007}.  In our
approach $P[n]$ is accurately calculated using the TFD theory
\cite{rossi2008} that takes into account nonlinear screening, exchange
and correlation contributions \cite{barlas2007,polini2008}, and can
handle the crossover plateau region where $P[n]$ has a bimodal
character that is not captured by the Gaussian and Lorentzian
approximations.  Only close to the Dirac point and for small values of
$r_s$ do the Gaussian and Lorentzian approximations give results that
are comparable to the ones obtained using $P[n]$ calculated with the
TFD theory, $P[n]^{TFD}$. The TFD approach allows us to go to higher
values of $r_s$ and lower $n_{\rm imp}$ thereby also providing a range
of validity for analytical theories.


\section{Results}
\label{sec_results}
\begin{figure}[ht]
 \begin{center}
  \includegraphics[width=8.00cm]{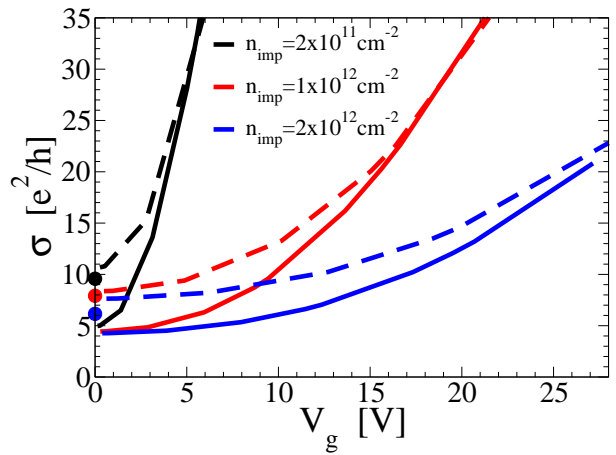}
  \caption{
           (Color online).
	   $\sigma$ as function of $V_g$ for 3 different values
	   of $n_{imp}$. The dashed (solid) lines are the results
	   without (with) exchange.
	   The dots at $V_g=0$ are the results presented in Ref.~\onlinecite{adam2007}
	   for the same values of $n_{imp}$.
          }
    \label{fig:s_vg}
 \end{center}
\end{figure}

\begin{figure}[ht]
 \begin{center}
  \includegraphics[width=8.00cm]{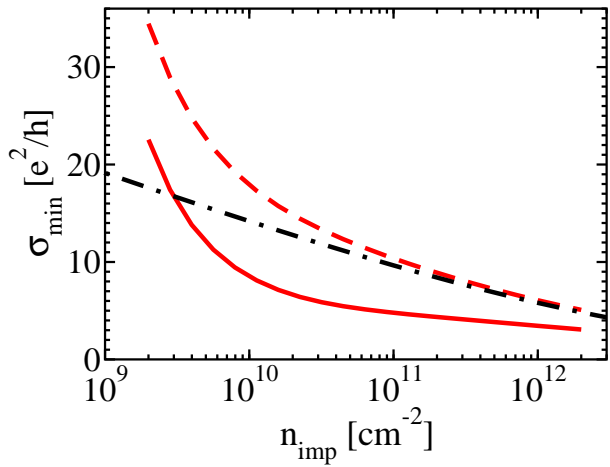}
  \caption{
           (Color online).
           $\sigma_{min}$ as a function of $n_{imp}$,
           $r_s=0.8$ and $d=1$~nm.
           Solid (dashed) line are the results with (without) exchange.
           For comparison the results obtained in Ref.~\onlinecite{adam2007} are also shown, dot-dashed line.
          }
    \label{fig:s_nimp}
 \end{center}
\end{figure}

\begin{figure}[ht]
 \begin{center}
  \includegraphics[width=8.00cm]{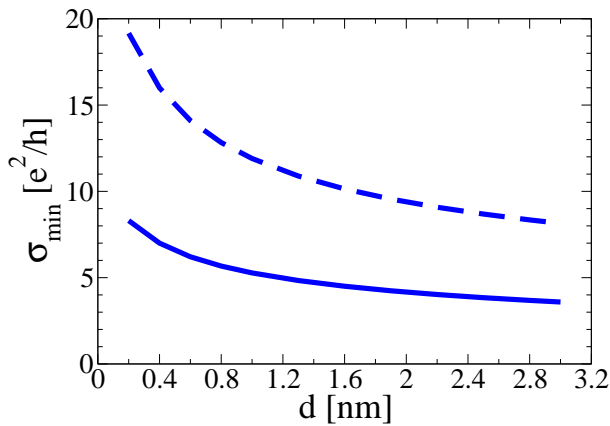}
  \caption{
           (Color online).
           Dependence of $\sigma_{min}$ on $d$ for 
           $n_{imp}=10^{11}$~cm$^{-2}$ and $r_s=0.8$.
           Solid (dashed) line are the results with (without) exchange.
          }
    \label{fig:s_d}
 \end{center}
\end{figure}

\begin{figure}[ht]
 \begin{center}
  \includegraphics*[width=8.00cm]{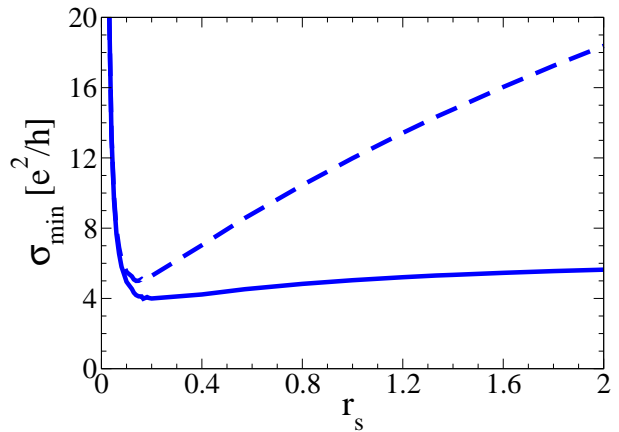}
  \caption{
           (Color online).
           $\sigma_{min}$ 
	   as a function of $r_s$ for $d=1$ and $n_{imp}=10^{11}$~cm$^{-2}$.
          }
    \label{fig:s_rs}
 \end{center}
\end{figure}

\begin{figure}[ht]
 \begin{center}
  \includegraphics*[width=8.00cm]{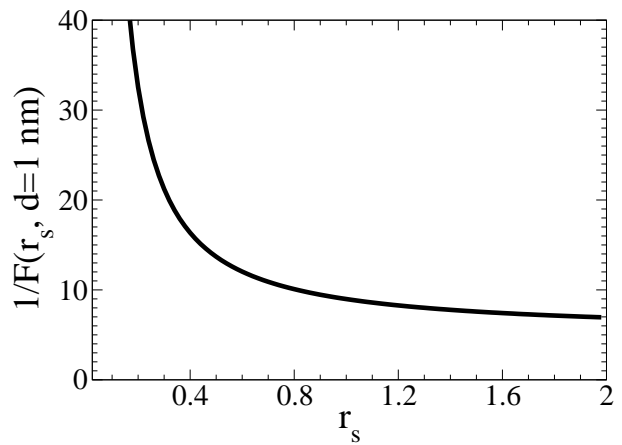}
  \caption{
           $1/F(r_s,d=1\;{\rm nm})$   
	   as a function of $r_s$.	   
          }
    \label{fig:Frs}
 \end{center}
\end{figure}

\begin{figure}[ht]
 \begin{center}
  \includegraphics*[width=8.00cm]{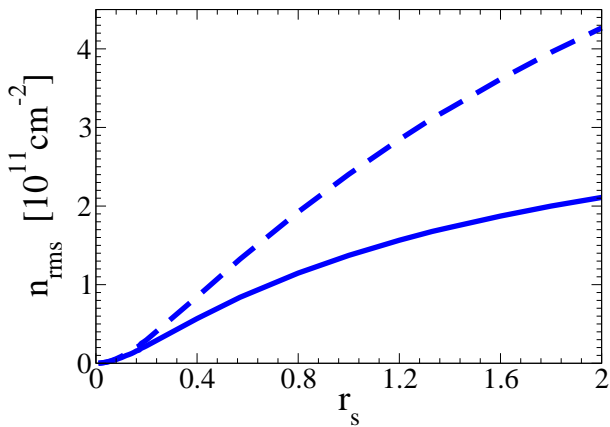}
  \caption{
           (Color online).
           $n_{rms}$,  
	   as a function of $r_s$ for $d=1$ and $n_{imp}=10^{11}$~cm$^{-2}$.
	   Solid (dashed) line are the results with (without) exchange.
          }
    \label{fig:nrms_rs}
 \end{center}
\end{figure}

In this section we  present the results for the conductivity
of disordered graphene obtained using the effective medium
theory developed in the previous section and the 
probability distribution, $P[n]^{TFD}$, obtained
using the TFD theory.  
Shown in  Fig.~\ref{fig:s_vg} is $\sigma(V_g)$.
We see that our theory correctly predicts a finite value of $\sigma$
at the Dirac point that is of the same order as that measured
experimentally.  At high gate voltages the linear scaling of $\sigma$
as a function of $V_g$ is recovered.  One important achievement of our
theory is the ability to correctly describe the crossover of $\sigma$
from its minimum, $\sigma_{min}$, at $V_g=0$, to its linear behavior
at high gate voltages.  Moreover our work shows the importance at low
gate voltages of the exchange term.
In Fig.~\ref{fig:s_nimp} we present the result for $\sigma_{min}$ as
function of $n_{imp}$. We see that our approach explains the
non-universality of $\sigma_{min}$ from sample to sample and that
exchange reduces the dependence of $\sigma_{min}$ on $n_{imp}$ for
impurity densities in the range $[10^{10}-10^{12}]$~cm$^{-2}$.
For $n_{imp}\lesssim$ $10^{10}$~cm$^{-2}$ $\sigma_{min}$ grows quite
rapidly as $n_{imp}$ is decreased.
The same qualitative dependence of $\sigma_{min}$ on $n_{imp}$
has been recently observed experimentally as shown 
in Fig.~5~(a) of Ref.~\onlinecite{chen2008}.
In Fig.~\ref{fig:s_d}, the dependence of $\sigma_{min}$ on $d$ is
shown for a given value of $n_{imp}$. We see that $\sigma_{min}$
depends weakly on $d$ and this dependence is even weaker if 
exchange-correlation effects are taken into account.
In Fig.~\ref{fig:s_rs} we show the results for $\sigma_{min}$
as a function of $r_s$.
The solid (dashed) line shows the values of $\sigma_{min}$ obtained
including (neglecting) exchange.  We see that $\sigma_{min}$ has a
non-monotonous behavior due to the fact that $r_s$ strongly affects
both the carrier density spatial distribution (by controlling the
strength of the disorder potential, screening and exchange); and the
scattering time $\tau$. From the RPA-Boltzmann
theory in presence of charged impurities we have that
$\tau = \sqrt{n}/(\pi F(r_s,d) v_F n_{imp})$, so that
$\tau$ depends on $r_s$ only through $F(r_s)$.
In Fig.~\ref{fig:Frs} we show $1/F(r _s)$ for $d=1$,
to show how $\tau$ decreases as $r_s$ increases.
On the other hand  as $r_s$ increases 
the width of $P[n]$, i.e. the number of carriers,
increases.  For small values of $r_s$ ($\lesssim 0.1$), as $r_s$
increases, the decrease of $\tau$ dominates and $\sigma_{min}$
decreases very rapidly.  In this regime the results without exchange
are qualitatively similar to the results with exchange.  For
$r_s\gtrsim 0.1$ the effect of $r_s$ on the amplitude of the spatial
density fluctuations becomes very important.  In Fig~\ref{fig:nrms_rs} 
the root mean square of the density
fluctuations is plotted as a function of $r_s$.  We see that in
presence of exchange $n_{rms}$ grows much more slowly than when
exchange is neglected. This is due to the fact that in graphene,
contrary to parabolic band materials, close to the Dirac point,
exchange penalizes density fluctuations. As a consequence for
$r_s\gtrsim 0.1$ the behavior of $\sigma_{min}$ is qualitatively
different when exchange is taken into account.  In
Ref.~\onlinecite{jang2008}, $r_s$ was varied from $0.56$ to $0.8$ and
$\sigma_{min}$ was found to remain constant in agreement with the
results shown in Fig.~\ref{fig:s_rs} in presence of exchange. We can
then draw the important conclusion that to understand, even at a
qualitative level, the transport properties of current graphene
samples exchange contributions must be taken into account.

\section{Discussion and Conclusions}
\label{sec_conclusions}
In this paper we have derived and justified the effective medium
theory for disordered bulk graphene. In particular we have considered
the case when the disorder is due to random charged impurities
in the graphene environment. Charge impurities have been shown
to be the most important source of disorder for current 
experiments on exfoliated graphene. There are however
other sources of disorder like ripples \cite{kim2008,fernando2007,brey2008,guinea2008}
and atomic defects \cite{lherbier2008} that
might play an important role in determining the
transport properties close to the Dirac point of graphene systems
other than bulk exfoliated graphene. For example
there is evidence that atomic defects might be
the dominant source of disorder in epitaxial graphene.
We should emphasize that our motivation was to calculate
the intrinsic conductivity of graphene close to the
Dirac point. For this reason we have neglected the effects
of contacts that have been shown to be important
\cite{lee2008, huard2008, giovannetti2008,golizadeh2009} but that do not modify the
value of the graphene intrinsic conductivity and in experiments
can be minimized by performing four-terminal measurements.
Our theory does not apply to graphene systems
with a band gap~\cite{adam2008} and is only valid in the diffusive limit.
To model the disorder we have assumed the impurities to
be randomly distributed in a 2D plane placed at a distance $d$
from the graphene. The only two input parameters that
enter the theory are the charge impurity density, $n_{imp}$,
and $d$. These two parameters have been accurately extracted
by comparing the RPA-Boltzmann theory and the experimental
results at high doping \cite{tan2007, chen2008}.
We have used the EMT developed here to study in details the 
transport properties of graphene close to the Dirac point,
where the physics is dominated by the 
strong density fluctuations induced by the disorder.
%
%
%
The parameters $n_{imp}$ and $d$ that enter the EMT 
should not be regarded as fitting parameters: 
to test our results for a given sample, $n_{imp}$ and $d$ 
should be extracted by comparing the value of the mobility
at high doping with the one provided by the RPA-Boltzmann
theory, and then used to obtain the EMT results at the
Dirac point. 
A key ingredient for the EMT is the accurate
calculation of the density probability distribution, $P(n)$,
close to the Dirac point taking into account non-linear
screening and exchange-correlation effects. 
We have used $P(n)$ provided by the TFD theory \cite{rossi2008}.
The TFD theory is justified for most of the current
experiment on exfoliated graphene and gives results
that, for single disorder realizations, are in agreement
with the DFT-LDA approach developed in Ref.~\onlinecite{polini2008}.
On the other hand due to its computational cost the
DFT-LDA does not allow the calculation of disorder averaged
quantities and therefore a statistical characterization
of the density profile.
The TFD-EMT approach that we have developed is valid when 
the inequalities \ceq{eq:ineq1}, \ceq{eq:ls2}, \ceq{eq:ineq3}
are satisfied. 
In these inequalities $n$
is the density inside the electron-hole puddles.
In the small regions, of size $\xi\approx 10$~nm, 
because of the high carrier density, inequality \ceq{eq:ineq1}
is always satisfied.
As we have
shown, most of the graphene is covered by large
puddles of size $L_p$ of the order of the system size, $L$,
and with density of the order of $n_{rms}$. As shown 
in Ref.~\onlinecite{rossi2008, adam2007} $n_{rms} \approx\eta n_{imp}$,
where $\eta$ is a number of order 1.
Using these results we find that \ceq{eq:ls2} and \ceq{eq:ineq3}
are equivalent apart from a factor of order 1 
and that the inequalities \ceq{eq:ineq1}, \ceq{eq:ineq3},
take the form:
\beq
 L_p \gg \frac{1}{\sqrt{\eta\pi n_{imp}}}; 
 \hspace{1.0cm}
 L_p \gg \frac{2\Gamma\sqrt{\eta}}{F(r_s, d)}\frac{1}{\sqrt{\pi n_{imp}}}
 \label{eq:ineqdp}
\enq
respectively. Given that for the cases of interest $1/F(r_s, d)\approx 10$, see
Fig.~\ref{fig:Frs}, the second inequality in \ceq{eq:ineqdp} is the
most stringent: the condition that the conductance across the PNJ must
be bigger than the conductance inside the puddles, or equivalently
the condition that the mean free path must be smaller than the puddles, 
sets the minimal size that the puddles must have for the EMT to be valid.
Considering that $L_p\sim L> 1\;\mu$m
and $n_{imp}\in[10^{10}-10^{12}]$~cm$^{-2}$, we see that for bulk
samples of exfoliated graphene the inequalities \ceq{eq:ineqdp} are
satisfied. We conclude that for sufficiently disordered and large
graphene samples the TFD-EMT approach is accurate to estimate the
graphene transport properties, this conclusion is consistent with the
results of Ref.~\onlinecite{adam2008d} in which the graphene
conductivity in presence of Gaussian disorder obtained
using a full quantum mechanical calculation was
found to be in agreement with the semiclassical Boltzmann one even at zero
doping provided the disorder is strong enough.

Using the TFD-EMT approach we find that at the Dirac point the theory
gives a finite minimum conductivity in very good agreement with that
observed experimentally \cite{tan2007,chen2008,jang2008}. In addition
the theory is able to describe the crossover of the conductivity from
its minimum value at the Dirac point to its linear behavior at high
doping.  We have also calculated the dependence of $\sigma_{min}$ on
$n_{imp}$ and found it to be in good agreement with recent
experimental results \cite{chen2008}.  We obtain the dependence of
$\sigma_{min}$ on $r_s$, finding that for $r_s\lesssim 0.1$,
$\sigma_{min}$ decreases very rapidly with $r_s$ but for $r_s\gtrsim
0.2$, exchange-correlation terms are qualitatively important and must
be included to correctly describe the minimum conductivity. This
result is in agreement with the experimental results presented in
Ref.~\onlinecite{jang2008} in which the minimum of conductivity is
found to be the same for $r_s=0.5$ and $r_s=0.8$.  By comparing our
results for $\sigma_{min}$ as a function of $n_{imp}$ and $r_s$ to the
experimental ones we find that exchange-correlation effects must be
included when evaluating the transport properties of graphene at the
Dirac point.  The reason is that exchange-correlation effects, close
to the Dirac point, at low $n_{imp}$ and high $r_s$, strongly suppress
the amplitude of the density fluctuations and therefore strongly
affect the transport properties.  The success of the EMT, coupled with
the microscopic disorder-induced graphene electronic structure
calculation, in obtaining the graphene conductivity near the Dirac
point indicates that this technique should be useful in calculating
many other properties of graphene in the theoretically difficult
inhomogeneity-dominated regime near the charge neutrality point.


\section{Acknowledgments}

We thank M. Fogler, M. Fuhrer, E. H. Hwang and R. Lutchyn for helpful discussions.
This work is supported by US-ONR and NSF-NRI.


\end{document}